# A DATA ACQUISITION SYSTEM FOR LONGITUDINAL BEAM PROPERTIES IN A RAPID CYCLING SYNCHROTRON*

J. Steimel[#], C.Y. Tan, FNAL, Batavia, IL 60510, U.S.A.

*Abstract*

A longitudinal beam properties, data acquisition system has been commissioned to operate in the Fermilab booster ring. This system captures real time information including beam synchronous phase, bunch length, and coupled bunch instability amplitudes as the beam is accelerated from 400 MeV to 8 GeV in 33 ms. The system uses an off-the-shelf Tektronix oscilloscope running Labview software and a synchronous pulse generator. This paper describes the hardware configuration and the software configuration used to optimize the data processing rate.

## INTRODUCTION

The Fermilab Booster is currently pushed to its intensity limit due to the high demand from different experiments for protons. This will continue into the era of the intensity frontier. While intensity is the most critical parameter, emittance must be controlled as well. Longitudinal emittance is an important parameter for optimizing the slip stacking process in the Main Injector. Longitudinal instabilities can spoil the emittance, so it is important to monitor bunch lengths and bunch phase oscillations in the Booster.

The Booster presents a special challenge when it comes to measuring beam properties. The acceleration ramp is short (33ms), and the RF frequency varies from 38–52.8 MHz. Standard instrumentation like spectrum analyzers and network analyzers cannot effectively measure beam properties, because they cannot track the changing frequencies. Early instrumentation involved a series of tracking oscillators, mixers, and diode detectors that would require vigilant calibration. There was a demand for a more robust means of gathering beam parameters.

## HARDWARE SETUP

A new system for acquiring longitudinal beam parameters was commissioned in the Booster. This system consists of a segmented memory scope that monitors a beam current monitor and a raw RF sum fanback signal from the cavities. The scope is triggered by a special mountain range trigger system.

### Segmented Memory Scope

There have traditionally been a trade-offs in sampled data acquisition: fast sample rates and memory depth. The re-trigger rate of the scopes was too slow to see more than one part of the Booster cycle in detail. However, these tradeoffs have been overcome by some modern scopes which have deep memories and a special triggering mode. If a single sweep does not fill its entire memory buffer, the scope can be re-triggered very quickly to store other sweeps in the memory buffer[1]. This quick re-triggering can go on until the entire memory buffer is full.

The Booster longitudinal data acquisition is done with a Tektronix TDS7154B Digital Phosphor Oscilloscope. The scope has a 20MS/s sample rate and 64MB deep memory. The current settings for the scope are 5MS/s with a 4us sweep time. This acquires slightly more than one Booster revolution period at the beam's lowest energy. The memory is deep enough to allow slightly more than 400 sweeps at the fast re-trigger rate, and the scope can be retriggered as fast as every other Booster revolution period. The scope is also a self contained Windows XP computer with an Ethernet connection.

### Mountain Range Trigger System

It would not be practical to always trigger the scope at its maximum trigger rate. The default mode of operation is a set of sweeps distributed through the Booster cycle. Triggering of the sweeps is accomplished with the mountain range triggering system[2]. The system counts RF cycles directly and can output triggers after a set number of counts. This keeps the system synchronized with the beam throughout the cycle. The system is also remotely programmable and can trigger on specified beam events dictated by the control system. To provide a set of sweeps distributed through the cycle, the system triggers the scope every 20 Booster revolution periods.

## DATA PROCESSING

Once the data from all the specified sweeps is collected, the data is analyzed locally. A Labview program, running on the scope, processes the data. The parameters that are calculated for each sweep are: fundamental frequency, RF voltage, beam synchronous phase, bunch width, and coupled bunch mode amplitudes.

### Calculating RF Frequency

An FFT is performed on each sweep as the first calculation in the data process. The sweep time dictates a frequency bin width of about 250kHz. This is too course for the required the actual frequency and amplitude of the RF is calculated by deconvolving the discrete FFT spectrum with a sinc function that has a half period equal to the bin width [3].

Table 1: Parameter Definitions

| Component | Definition | Value |
|---|---|---|
| $f_{bin}$ | Frequency spacing between FFT points | Sample rate/# of points |
| FFT[i] | FFT Array of complex amplitudes | |
| $k$ | FFT Array index of maximum value | Max(FFT[i])=FFT[k] |
| $A_k e^{i\vartheta_k}$ | Complex amplitude of max(FFT) | FFT[k] |
| $A_{k\pm} e^{i\vartheta_{k\pm}}$ | Complex amplitude of values closest to max(FFT) | FFT[k±1] |
| $f_{rf}$ | Beam RF frequency | |
| $r$ | Real FFT array index of $f_{rf}$ | $f_{rf}/f_{bin}$ |
| $A_{rf} e^{i\vartheta_{rf}}$ | Complex amplitude of RF frequency | FFT[r] |
| $A_{3rf} e^{i\vartheta_{3rf}}$ | Complex amplitude of 3rd RF frequency harmonic | FFT[3r] |
| $A_3 e^{i\vartheta_3}$ | Complex amplitude of FFT[i] closest to 3rd RF frequency harmonic | FFT[round(3r)] |
| $A_{nrev} e^{i\vartheta_{nrev}}$ | Complex amplitude revolution frequency | FFT[(r+r*n/84)] |
| $A_n e^{i\vartheta_n}$ | Complex amplitude of FFT[i] closest to revolution frequency | FFT[round(r+r*n/84)] |

The simple deconvolution involves finding the peak amplitude in the FFT (assumed fundamental frequency) and the largest signal within a bin width of the peak. The formula for calculating the real array index that represents the actual fundamental frequency is:

$$r = k - \frac{\frac{A_{k+}}{A_k}\cos(\theta_{k+}-\theta_k)}{1-\frac{A_{k+}}{A_k}\cos(\theta_{k+}-\theta_k)} \quad (1)$$

or

$$r = k - \frac{\frac{A_k}{A_{k-}}\cos(\theta_k-\theta_{k-})}{1-\frac{A_k}{A_{k-}}\cos(\theta_k-\theta_{k-})} - 1 \quad (2)$$

where the adjacent bin with the largest amplitude has been chosen to reduce errors. See Table 1 for parameter definitions.

The complex amplitude of the RF frequency becomes:

$$A_{rf} = \frac{A_k(r-k)}{\sin(\pi(r-k))} \quad (3)$$

$$\theta_{rf} = \theta_1 + \pi(r-k) \quad (4)$$

This calculation is repeated for the RF sum and beam current signals. This gives us the RF amplitude seen by the beam and a phase difference between the beam and RF. Compensating for the electrical delay between the beam and RF signals gives the beam synchronous phase shown in Figure 1.

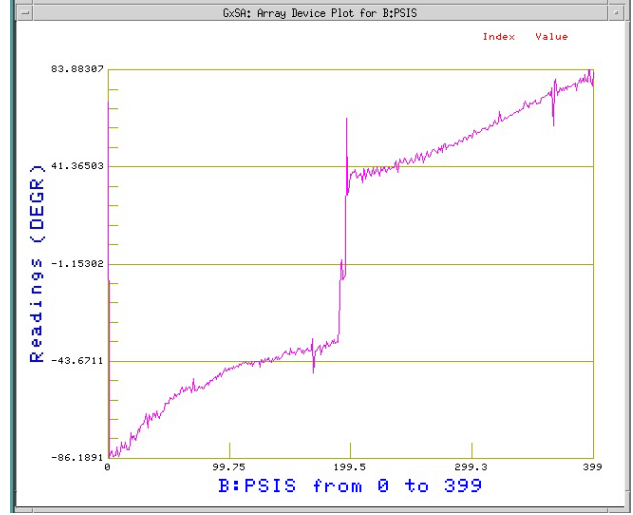

**Figure 1: Plot of beam synchronous phase though the Booster acceleration cycle. Jump in phase is due to transition crossing.**

*Calculating Bunch Width*

To calculate the longitudinal bunch width, the approximation that the beam is Gaussian has been made. For purely Gaussian beam there is a formula for bunch width which is based on the ratio of amplitudes between the first and third fundamental RF harmonics[4].

$$\sigma_{rms} = \frac{0.108}{f_{rf}}\sqrt{20\log(\frac{A_{rf}}{A_{3rf}})} \quad (5)$$

The non-integer index value for $A_{3rf}$ is exactly three times the non-integer index value for the fundamental frequency. The actual amplitude can be derived from the amplitude of the closest integer index to the third harmonic. However, because of the short sweep time and high relative amplitude of the fundamental RF in the beam current spectrum, the fundamental RF has significant bleed through into other frequency bins. The formula for the fundamental bleed through is given by:

$$\Delta A_{rf} = A_{rf} e^{i\theta_{rf}} \frac{\sin(\pi(r-i))}{r-i} \quad (6)$$

where $i$ is the index of the bin to be analyzed. The formula for the third harmonic amplitude is:

$$A_{3rf} = \frac{(A_3 - \Delta A_{rf})(3r - \text{round}(3r))}{\sin(\pi(3r - \text{round}(3r)))} \quad (7)$$

where $A_3$ is the amplitude of the integer FFT index closest to the third harmonic.

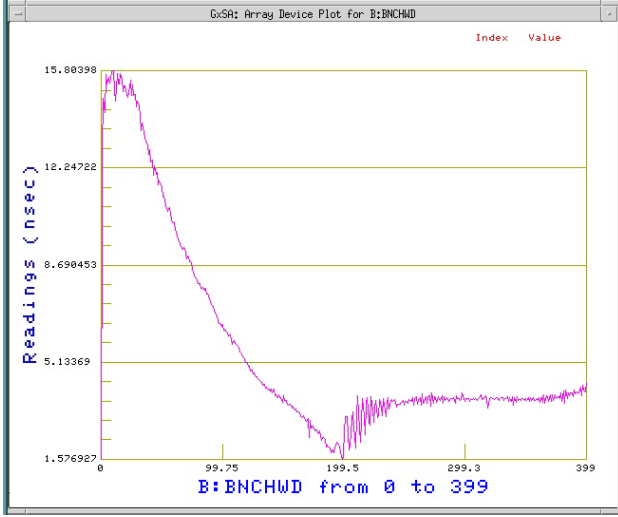

**Figure 2:** Plot of beam bunch width through the Booster acceleration cycle.

### Calculating Coupled Bunch Mode Amplitudes

The sweep time of the beam signal is much too short to discern synchrotron lines in the FFT. The only way to monitor coupled bunch mode amplitudes is to extract the information from the revolution harmonics. Revolution frequency amplitudes are calculated in much the same way as the third harmonic amplitude is calculated. First, the desired revolution frequency index is derived from the RF frequency index.

$$f_{nrev} = \left(1 + \frac{n}{h}\right) f_{rf} \quad (8)$$

where $n$ is the revolution harmonic and h is the harmonic number of the Booster which is 84. Thus, the formula for the revolution frequency amplitude is:

$$A_{nrev} = \frac{(A_n - \Delta A_{rf})(r(1+\frac{n}{h}) - \text{round}(r(1+\frac{n}{h})))}{\sin(\pi(r(1+\frac{n}{h}) - \text{round}(r(1+\frac{n}{h}))))} \quad (9)$$

where $A_n$ is the amplitude of the integer FFT index closest to the revolution harmonic.

The intensity of the revolution harmonic is typically dominated by bunch population, acting like an amplitude modulation on the fundamental RF. Coupled bunch instabilities are pure phase modulation of the fundamental RF. If we assume that all components of phase modulation of the beam are due to coupled bunch modes, the coupled bunch mode amplitude is:

$$a_n \sim |A_{-nrev} - A^*_{nrev}| \quad (10)$$

This formula removes the component of the revolution lines that is due to amplitude modulation.

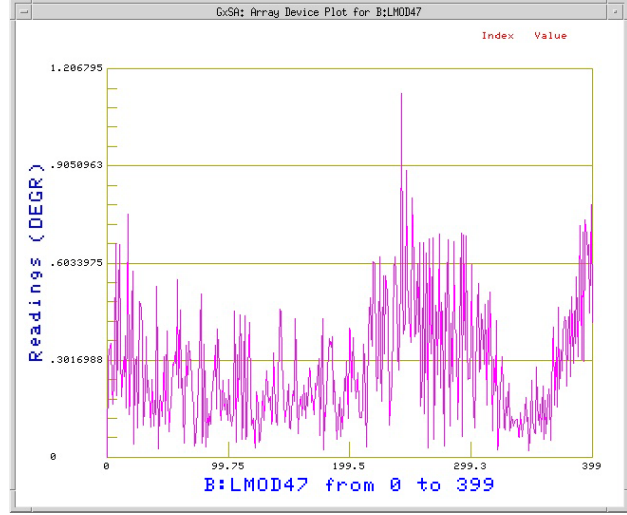

**Figure 3:** Plot of the amplitude of coupled bunch mode 47 through the Booster cycle.

## PERFORMANCE

One of the key parameters in the design of the hardware layout and processing is the overall retrigger rate. Machine operators need feedback on beam cycles quickly to speed in tuning operations. However, the system needs to process 64MB of data. This process is expedited by the fact that the data is processed on the scope itself, so there is no need to transfer the data across the network. The current retrigger rate is about once every seven seconds and the majority of time is spent transferring data from the scope to the scope computer.